\pdfoutput=1
\documentclass[a4paper]{article}

\usepackage{INTERSPEECH2022}

\usepackage{xcolor}

\newcommand{\sstitle}[1]{\smallskip\noindent\textbf{#1.\/}}

\newcommand{\eg}{e.\,g.,\ }
\newcommand{\ie}{i.\,e.,\ }

\usepackage{hyperref}
\usepackage{booktabs}
\usepackage{paralist}
\usepackage{float}
\usepackage{balance}
\usepackage{caption} 
\usepackage{subcaption}
\usepackage{graphicx}

\usepackage[utf8]{inputenc}
\usepackage[activate]{microtype}

\title{Example-based Explanations with Adversarial Attacks \\ for Respiratory Sound Analysis}

\name{Yi Chang$^{1,*}$, Zhao Ren$^{2,*}$, Thanh Tam Nguyen$^3$, Wolfgang Nejdl$^2$, Björn W. Schuller$^{1,4}$
\thanks{* Y. Chang and Z. Ren contribute equally. T. T. Nguyen is the corresponding author. This research was partially funded by the Federal Ministry of Education and Research (BMBF), Germany under the project LeibnizKILabor with grant No.\,01DD20003. The code is released at \url{https://github.com/glam-imperial/SoundPrototypeCriticism}.}}
\address{
\small
  $^1$GLAM -- Group on Language, Audio, \& Music, Imperial College London, United Kingdom\\
  $^2$L3S Research Center, Leibniz Universit\"at Hannover, Germany\\
  $^3$Griffith University, Australia\\
  $^4$EIHW -- Chair of Embedded Intelligence for Health Care and Wellbeing, University of Augsburg, Germany
  }
\email{y.chang20@imperial.ac.uk,zren@l3s.de,t.nguyen19@griffith.edu.au,nejdl@l3s.de,schuller@ieee.org}
\begin{document}

\maketitle
\begin{abstract}
 
Respiratory sound classification is an important tool for remote screening of respiratory-related diseases such as pneumonia, asthma, and COVID-19. To facilitate the interpretability of classification results, especially ones based on deep learning, many explanation methods have been proposed using 
%examples and 
prototypes.
% While the success of deep learning boosts the performance of respiratory sound classification recently, existing approaches suffer from the non-explainable issues, i.e. the predictions of the models are difficult to interpret for human experts. 
However, existing explanation techniques often assume that the data is non-biased and the prediction results can be explained by a set of prototypical examples. In this work, we develop a unified example-based explanation method for selecting both representative data (prototypes) and outliers (criticisms). In particular, we propose a novel application of adversarial attacks to generate an explanation spectrum of data instances via an iterative fast gradient sign method. Such unified explanation can avoid over-generalisation and bias by allowing human experts to assess the model mistakes case by case. We performed a wide range of quantitative and qualitative evaluations to show that our approach 
generates effective and understandable explanation
%outperforms existing explanation baselines 
and is robust with many deep learning models. 
%Our source code is available at \url{https://github.com/leibniz-future-lab/SoundPrototypeCriticism}.

\end{abstract}


\section{Introduction}
Respiratory sound classification plays an important role in today's diagnosis systems to assist physicians in identifying adventitious sounds~\cite{pramono2017automatic}. While respiratory diseases such as COVID-19, bronchial asthma, and chronic obstructive pulmonary disease are affecting more and more the world population~\cite{yang20e_interspeech}, such computer-aided auscultation of respiratory sounds provides a remote and non-invasive instrument for early screening of the diseases. Owing to its promising prospect, respiratory classification has been studied intensively~\cite{chambres2018automatic,ramanarayanan2020toward,rasskazova2019temporal}. Especially, the success of deep neural networks (DNNs) in various application domains also boosts recent studies of respiratory sounds with better predictive accuracy~\cite{pramono2017automatic,yang20e_interspeech}.

However, these advances have also introduced increasing complex and black-box models that are not explainable by nature, \ie their decision boundaries are difficult to understand~\cite{li2022explainable}. As a result, it is difficult for healthcare practitioners to fully trust the predictions if no explanation is available, especially when many respiratory sound classification results still have modest performance (\eg the average score of around 50.16\,\% on the ICBHI 2017 dataset~\cite{rocha2019open,yang20e_interspeech}). 
%BS: unclear - do you mean 50% of that score or is 50% the score?
Existing works tried to mitigate this problem with data augmentation~\cite{song2021contrastive} to feed more data to DNNs. Nevertheless, the interpretability of a model is crucial in high-stake domains such as healthcare, where mistakes can cause significant damage~\cite{ahmad2018interpretable,du19_interspeech}.

%most of respiratory classification results still contain many mistakes and biases, which leads to modest performance (e.g. XX\% accuracy~\cite{}). This could be explained by the fact that respiratory datasets are still in development and not quite large enough to leverage the full potential of deep neural networks. 

Despite many recent advances in explainable artificial intelligence (AI) to mitigate mistakes~\cite{abderrazek2020towards,schiller19_interspeech,nauta2021neural}, there are still enormous challenges for explaining respiratory sound classification. Most existing explainable methods focused on attention mechanisms (\eg identifying parts of the input that most contributed to the final model decision)~\cite{ren2020caanet,ren2022prototype}. Other works focused on example-based explanations using prototypes, which are data instances representative of a target class~\cite{gurumoorthy2019efficient,ming2019interpretable}. Unlike attention mechanisms that do not provide actionable insights of the models, example-based explanations facilitate cognitive human understanding, in particular case-based reasoning~\cite{stock2018convnets}, as well as 
%BS: there is a word missing here?
%hold 
vast potential to improve the classification quality via nearest neighbour classifiers~\cite{ren2022prototype,kim2016examples}.

However, existing example-based explainable methods do not consider bias in data (as is often the case in real-world data). In fact, the distribution of a model decision cannot be represented by a set of prototypical examples, but also criticisms -- data instances sampled from regions of the input space not well captured by the model. These criticism examples often lie close to the decision boundary of the same target class and often represent model mistakes (\eg false positives) or out-of-distribution data. Indeed, including criticisms into explanations can avoid over-generalisation and bias by allowing human experts to assess the misclassified examples and outliers~\cite{kim2016examples,stock2018convnets}.

An adversarial attack is a common tool to uncover model mistakes and biases by injecting small adversarial perturbations into existing inputs. Such perturbed inputs, called adversarial examples, are indistinguishable from original inputs by a human, yet,they are capable of fooling the model to change the target class~\cite{ren2020generating,ren20_interspeech}. Motivated by adversarial attacks, we develop a unified solution for example-based explanations using prototypes and criticisms. Instead of using adversarial perturbations to change the target class, we consider a novel application of the adversarial perturbations to generate a spectrum of data instances that include both prototypes and criticisms simultaneously. In particular, we propose an iterative fast gradient sign method (IFGSM) for generating perturbations, which offer a natural way of selecting prototypes and criticisms based on the number of steps of IFGSM.

%necessary to change the representation for a given example as much as possible while still retaining the original decision. 
%\todo{Tam: please rewrite the technical description of our approach if necessary}.

Our work relates closely to existing works on prototype learning and criticism learning such as MMD-critic~\cite{kim2016examples} and ProtoDash~\cite{gurumoorthy2019efficient}. These works started selecting a set of prototypes first, then separate them into prototypes and criticisms by solving a submodular optimisation problem on data distribution. However, these methods are model-agnostic (explanations are completely independent of the model) or only indirectly capture the discriminative nature of a model via a hidden kernel-based representation of the examples. Unlike these works, we argue that adversarial attacks can be used to unify example-based explanations, \ie prototypes at one end and criticisms at the other end of the explanation spectrum generated by the IFGSM.

To the best of our knowledge, this is a novel application of adversarial attacks for explainable respiratory sound classification. We propose an interpretable and steerable explanation process for any type of DNNs. In doing so, we overcome the challenges of interpretability in audio data, which often exhibit high-order structures in temporal, spatial, and spectral dimensions. Especially, our approach in unifying prototypes and criticisms via adversarial attacks would benefit users in many ways: (i) the selected prototypes and criticisms can uncover new cases or outliers about the diseases, and (ii) they can be further analysed by post-hoc analysis such as attention tensor learning.

%the selected prototypes and criticisms can be further analyzed by added attention matrix learning, (ii) users both understand the typical cases and extreme cases of model decision, and (iv) a criticism can be a new case so that the experts can understand more about the diseases.

%(ii) it is easier to visual compare a classified audio sample and the selected examples

%\begin{figure}[!h]
%	\centering
%	\includegraphics[width=1.0\linewidth]{figures/respiratory.png}
%	\caption{Dummy}
%	\label{fig:dummy}
%\end{figure}

%\sstitle{Respiratory Sound Analysis}
%\edit{Respiratory sounds are commonly used to diagnose obstructive or restrictive lung diseases~\cite{li2022explainable}. By examining respiratory sounds during auscultation, the medical practitioners can identify adventitious sounds (e.g., crackles or wheezes) during the respiratory cycle. For example, crackles are the earliest sign of idiopathic pulmonary fibrosis and wheezes are usually related to COPD and asthma [5]. How- ever, the traditional way to detect abnormalities in lung sounds by using a stethoscope could be affected by various factors such as the environmental noise, hearing fatigue, and lack of experience among junior doctors. In addition, traditional stethoscopes are not user-friendly when doctors have to wear full Personal Protective Equipment (PPE) in highly infectious environment.}
%
%\cite{yang20e_interspeech}

\sstitle{Related Work}
Most existing approaches to respiratory sound classification neglect the question \emph{why} certain patients have been classified as a target class. Although there exists many interpretable methods such as regression weights and attention maps~\cite{abderrazek2020towards,gao20_interspeech}, they are difficult to validate in sound data, which often exhibit high-order structures in temporal, spatial, and spectral domains. We argue that explanations shall be based on a set of evidential examples, which enable human experts to compare real examples and generalise the problem properties. Some works tried to do so with prototype layers~\cite{thienpondt20_interspeech,ren2022prototype,yang2018robust}, which, however, are synthetic and biased, as the model is forced to focus more on typical examples and ignore extreme cases such as outliers and under-sampled data. 
%Prototype learning is an emerging interpretable machine learning paradigm that explains and classifies at the same time by comparing the inputs to a few prototypes, generated in an additional prototypical layer in the middle of a DNN~\cite{thienpondt20_interspeech,ren2022prototype,yang2018robust}. However, doing so might increase bias in the data, as the model is forced to focus more on typical examples and ignore extreme cases such as outliers and under-sampled data. 
%Moreover, the prototypes are often generated, and thus synthetic, making it more difficult for human experts to capture real-world properties. 
Our work is a first attempt to unify example-based explanation for respiratory sound classification by creating an explanation spectrum of real examples to cover normal and abnormal characteristics of data.

\section{Methodology}

%To introduce our approach of example-based explanation, 
Let us define a dataset $\mathcal{D}=\{(\mathcal{X}, y)\}_{i=1}^n$, where $\mathcal{X}$ is the features, $y$ denotes the labels, and $n$ is the number of data samples.
In the following, we will firstly give the definitions of prototypes and criticisms, and then explain the adversarial attacks that are used in our study. Finally, the whole process of our example-based explanation will be described.

\subsection{Explanation Spectrum}
Based on the dataset $\mathcal{D}$, a set of prototypes and criticisms will be searched out to represent the distribution of the model decision.

\sstitle{Prototypes}
A strong DNN model often has small intra-class variations and relative large inter-class variations in a classification task~\cite{he2018triplet}. Ideally, the high-level representations learnt by a DNN could be split into $N$ groups according to $N$ classes. The data sample at the centre of each group can be considered as the most representative example for the corresponding class. However, it is challenging to have such an ideal data distribution on real-world data due to a range of reasons, such as noise. Therefore, the high-level representations may be learnt into more than $N$ groups. In this context, \emph{prototypes} are the most representative examples of each group. 

\sstitle{Criticisms}
Despite multiple groups for each class, the high-level representations of real-world data often have outliers~\cite{stock2018convnets}. Although the outliers have only a few samples, they should be still correctly predicted by the model. Nevertheless, it is difficult to search or learn prototypes to represent only a small part of data in each class. In this regard, we call the data samples close to these outliers \emph{criticisms}. Criticisms can be the outliers themselves or generated by the DNN model. In our study, both prototypes and criticisms are searched data examples for example-based explanation.

\subsection{Adversarial attacks}
DNNs have shown their vulnerability to adversarial attacks on acoustic tasks in our prior studies~\cite{ren2020generating,ren20_interspeech}. Due to the data distribution of prototypes and criticisms, we have an assumption that prototypes are the most difficult to be attacked, as they are close to the centre of the class groups; and the criticisms are very easy to be attacked, as they are the outliers. In this study, we search the prototypes and criticisms by the perturbations generated by white-box attack IFGSM~\cite{ren2020generating} which is stronger than FGSM due to its $I$ iterative steps. FGSM calculates the perturbations based on the gradient of loss function. For a data sample $\mathcal{X}_i$, the perturbation is generated by
\begin{equation}
    \mathcal{X}_{i}^{per} = \epsilon\ \mbox{sign}(\nabla_{\mathcal{X}_i}\mathcal{L}(\theta,\mathcal{X}_i,y_i')),
    \vspace{-5pt}
\end{equation}
where $\epsilon$ is a coefficient that controls the difference between the perturbation and the original data, $\mathcal{L}$ is the loss function, $\theta$ stands for the model parameters, and $y_i'$ is the predicted label of $\mathcal{X}_i$. Finally, the adversarial sample is calculated by $\mathcal{X}_i^{adv}=\mathcal{X}_i+\mathcal{X}_{i}^{per}$. 
Another benefit of using adversarial attacks is that the deeper the model is, the easier it is to attack~\cite{ren2020generating}.

\subsection{Example-based explanation process}

\begin{figure}[t]
    \centering
    \includegraphics[trim={0 7.9cm 6cm 4.6cm},clip,width=1.0\linewidth]{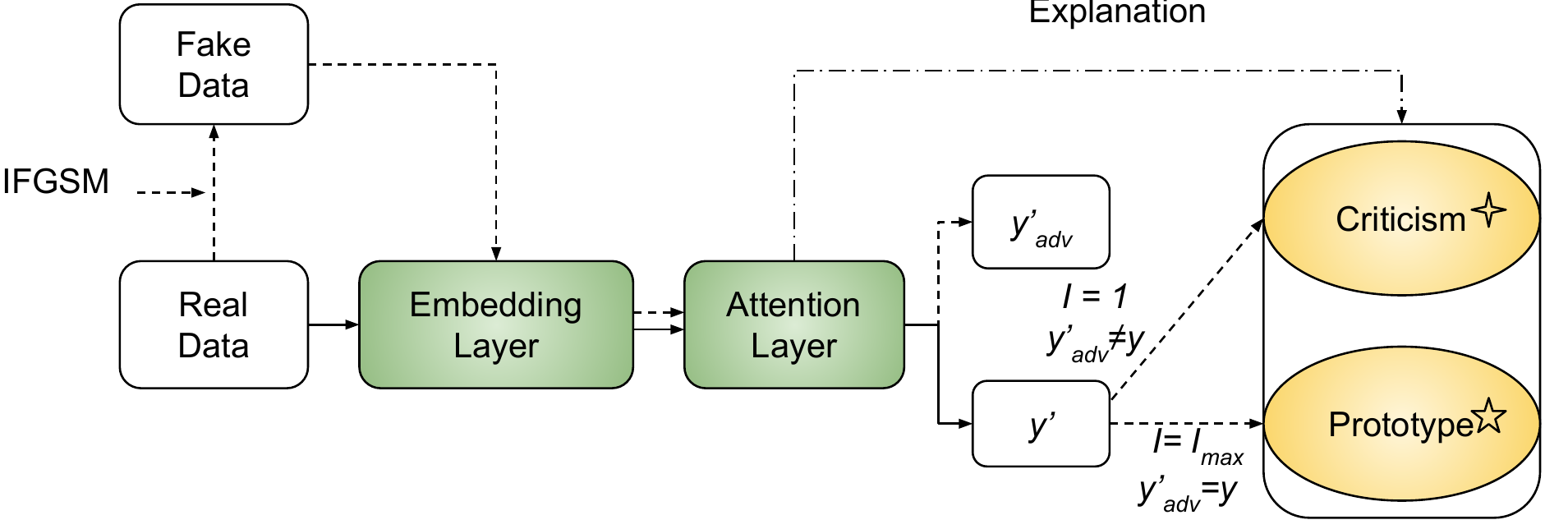}
    \caption{%The proposed example-based explanation process. (a) 
    The explanation pipeline with adversarial attacks. The solid lines are the data flow of real data, the dash lines are for adversarial (\ie fake) data, and the dash line with dots is the explanation procedure with attention.
    %(b) The distribution of prototypes and criticisms. 
    %\zr{Please make it better if you have better ideas: https://docs.google.com/drawings/d/1y49Iq5Ahxs9IxJW96oKMVhPhW2enGHHG98HiN7YpLo4/edit?usp=sharing}
    }
    %\vspace{-15pt}
    \label{fig:framework}
     \vspace{-15pt}
\end{figure}

Inspired by the study of~\cite{stock2018convnets}, adversarial attacks (\ie IFGSM) can be an efficient alternative to MMD-critic~\cite{kim2016examples} for prototype and criticism selection. \autoref{fig:framework} shows our approach of selecting prototypes and criticisms. Specifically, since prototypes are the most representative examples, prototypes should be still correctly classified (\ie $y'_{adv}=y$) after a certain number of maximum steps $I_{max}$ of FGSM attack. Similarly, because the criticisms are those samples that are not well captured by the model, they should be very vulnerable to the IFGSM attack. Therefore, criticisms tend to be misclassified (\ie $y'_{adv}\ne y$) after just one step or very few steps of IFGSM. Notably, the prototypes and criticisms are selected from the real data, since adversarial data may lie in a different data distribution, especially for criticisms.

To further verify and explain the selected prototypes and criticisms, it is essential to know which parts of these samples can effectively represent the corresponding distributions. We tackle this challenge by training DNNs with an embedding layer with dilation and an attention layer. Dilated kernels in the embedding layer focus on preserving the size of feature maps, and the attention layer aims to learn the potential contribution of each unit in the prototypes and criticisms~\cite{ren2018attention,ren2019attention,ren2020caanet}.

% \zr{This description could be more clear to describe figure 1. more importantly, more mathmatical representations should be involved.}
% Inspired by the study of~\cite{stock2018convnets}, adversarial attacks (\ie IFGSM) is an efficient alternative to MMD-critic~\cite{kim2016examples} for prototype and criticism selection. Specifically, since prototypes are the most representative examples, prototypes should be still correctly classified after a certain number of maximum steps $M$ of IFGSM attack. Similarly, because the criticisms are those samples that are not well captured by the model, they should be very vulnerable to the IFGSM attack. Therefore, criticisms tend to be misclassified after just one step or very few steps of IFGSM. 

\section{Experiments and Results}

\subsection{Experimental Settings}

\sstitle{Data}
To verify our proposed approach, our study is based on the Scientific Challenge database released at the International Conference on Biomedical and Health Informatics (ICBHI) 2017~\cite{rocha2019open}, which is the largest publicly available acoustic database for respiratory sound classification. From seven chest locations 
%(\ie trachea, anterior left, anterior right, posterior left, posterior right, lateral left, and lateral right) 
of $126$ participants, $920$ audio waves were recorded with four devices, \ie one microphone and three stethoscopes. From all recordings, totally $6\,898$ respiratory cycles were derived. Each respiratory cycle was annotated with one of the four classes, \ie normal, crackle, wheeze, and both (crackle $+$ wheeze). In the ICBHI challenge, the database was split into a training set ($60\,\%$) and a test set ($40\,\%$). Similar to our prior study~\cite{ren2022prototype}, the training set is further split into a train set ($70\,\%$) and a validation set ($30\,\%$) for optimising the model hyperparameters. Notably, the split procedure is subject-independent to avoid the data from the same person appear in both train and validation set. The data distribution of the database on the four classes and the three datasets is described in~\autoref{tab:datasets}.

%The Scientific Challenge database released at the International Conference on Biomedical and Health Informatics (ICBHI) 2017~\cite{rocha2019open} is the largest publicly available collection of audio samples for respiratory sound classification. Totally $920$ audio recordings were collected from seven chest locations (\ie trachea, anterior left, anterior right, posterior left, posterior right, lateral left, and lateral right) of $126$ participants with four devices (\ie one microphone and three stethoscopes). The audio recordings have different sampling rates: $4$\, kHz, $10$\, kHz, and $44.1$\,kHz. All recordings derive $6\,898$ respiratory cycles, each of which was annotated with one of the four classes: \emph{normal}, \emph{crackle}, \emph{wheeze}, and \emph{both}, \ie \emph{crackle $+$ wheeze}. The database was split into a training set ($60\,\%$) and a test set ($40\,\%$) for the competition. 
%To optimise the model hyperparameters, we further divide the training set into two subject-independent data sets: a train set ($70\,\%$) and a development set ($30\,\%$) (see \autoref{tab:datasets}). 

\begin{table}[!h]
    \centering
    \footnotesize
%    \vspace{-5pt}
    \caption{The data distribution of the ICBHI database. }
    \label{tab:datasets}    
    \vspace{-10pt}
    %\scalebox{0.85}{
    \begin{tabular}{l|p{1cm}p{1cm}p{1cm}p{1cm}}
    \toprule
         \#& \textbf{Train} & \textbf{Devel} & \textbf{Test} & $\bm{\sum}$ \\
         \hline
         \textbf{Normal}& 1\,513 & \ \ \,550 & 1\,579 & 3\,642\\
         \textbf{Crackle}&\ \ \,616 & \ \ \,599 & \ \ \,649 & 1\,864 \\
         \textbf{Wheeze} & \ \ \,281 & \ \ \,220 & \ \ \,385 & \ \ \,886 \\
         \textbf{Both} & \ \ \,131 & \ \ \,232 & \ \ \,143 & \ \ \,506 \\
         \hline
         $\bm{\sum}$ & 2\,541 & 1\,601 & 2\,756 & 6\,898 \\
         \bottomrule
    \end{tabular}
    %}
%    \vspace{-10pt}
\end{table}

% \begin{table}[]
%     \centering
%     \scriptsize
%     \vspace{-5pt}
%     \caption{Distribution of the splitted train/development sets and the official test set in the ICBHI database. }
%     \label{tab:datasets}    
%     \vspace{-10pt}
%     \begin{tabular}{l|p{1cm}p{1cm}p{1cm}p{1cm}p{1cm}}
%     \toprule
%          \#& \textbf{Normal} & \textbf{Crackle} & \textbf{Wheeze} & \textbf{Both} & $\bm{\sum}$ \\
%          \hline
%          \textbf{Train}& 1\,513 & \ \, 616 & 281 & 131 & 2\,541\\
%          \textbf{Devel}& \ \, 550& \ \, 599 & 220 & 232 & 1\,601\\
%          \textbf{Test} & 1\,579 & \ \, 649 & 385 & 143 & 2\,756 \\
%          \hline
%          $\bm{\sum}$ & 3\,642 & 1\,864 & 886 & 506 & 6\,898 \\
%          \bottomrule
%     \end{tabular}
%     \vspace{-10pt}
% \end{table}

\sstitle{Evaluation Metrics}
We report the \emph{unweighted average recall} (UAR) as the generic classification benchmark instead of \emph{accuracy}, as UAR can provide fairer evaluation of the models over the four classes than \emph{accuracy}~\cite{song2021contrastive,ren2020generating}
%BS: added:
in case of imbalance. 
It is also common to distinguish abnormal audio samples (\ie crackles, wheezes, and both) from normal cases. Therefore, the following standard benchmarks are officially used in the ICBHI challenge~\cite{rocha2019open}: \emph{sensitivity} (SE) -- the number of true abnormal cases over the total number of abnormal cases, \emph{specificity} (SP) -- the ratio of true normal cases over normal cases, and \emph{average score} (AS) -- the average of SE and SP.

% \sstitle{Baselines}
% We compare our approach with several prototype and criticism selection algorithms:
% \begin{compactitem}
% \item Maximum mean discrepancy (MMD)-critic~\cite{kim2016examples} focused on example-based explanations by searching for prototypes and criticisms. Prototypes were searched by minimizing the square of MMD between the dataset containing prototypes and the other datasets, and the criticisms were selected from the data samples which lead to the maximum of the similarity between the dataset and the prototypes. Compared to MMD, our approach effectively use adversarial attacks to select prototypes and criticisms rather than apply hidden kernel-based representations. 
% \item ProtoDash~\cite{gurumoorthy2019efficient} proposed a single coherent framework for selecting prototypes and criticisms, and provided non-negative weights for selected prototypes to indicate their importance. Similar to MMD-critic, ProtoDash also used a kernel function in the process.  
% \item Adversarial attacks~\cite{stock2018convnets} were used to search for prototypes and criticisms, and employed local interpretable model-agnostic explanations (LIME) to explain the model predictions on prototypes and criticisms. Compared to this work, our approach involves the local expolanation in the model training procedure.
% \end{compactitem}

%\sstitle{Implementation Details}

\sstitle{Preprocessing} 
All audio recordings are re-sampled into $4$\,kHz. Since there are confounding noises in most files of the dataset (\eg handling noise, speech)~\cite{rocha2019open}, we apply a fifth 
%BS: added:
order 
butterworth bandpass filter as the denoising technique. Further, all respiratory cycles with various durations are unified into $4$\,s. Specifically, $4$\,s of audio signals are randomly chosen in the training procedure for better flexibility, whereas in the testing process, such length of audio signals are selected in the middle of each respiratory cycle for better performance. With the selected audios, we extract the log Mel spectrograms with a sliding window size of $256$, a hop length of $128$, and $128$ Mel bins.

\sstitle{Model Architechture} 
In the CNN8 encoder, there are four convolutional blocks with output channel numbers of $64$, $128$, $256$, and $512$, respectively. Each of the convolutional blocks is composed of two convolutional layers with the kernel size $3 \times 3$, followed by a local max pooling layer with a kernel size of $2 \times 2$. For fair comparison, there are also four convolutional blocks in the ResNet encoder. Similarly, the output channel numbers of convolutional blocks are $64$, $128$, $256$, and $512$, each of which applies the `shortcut connections' to add the identity mapping with the outputs of two stacked $3 \times 3$ convolutional layers~\cite{chang2021covnet}. For the classification, we either apply a global max pooling layer followed by an FC layer or a global attention pooling layer to learn the contribution of each time-frequency bin. For the dilated CNN8 and ResNet, the dilation rates applied for each convolutional block are $1$, $2$, $4$, and $8$, 
where each convolutional layer shares the same dilation rate, besides the one for the `shortcut connections'~\cite{chang2021covnet}.
%\yc{Zhao, do we need to describe the dilation and attention mechanism somewhere before this section? Like the equation 5 and 6 in the https://arxiv.org/pdf/2011.09299.pdf}

\sstitle{Model training} 
During training, we utilise the `Adam' optimiser with an initial learning rate of $0.001$ and set the batch size as 16. Specifically, the learning rate is decayed by a factor of $0.9$ at every $200$-th iterations for stabilisation. All training processes are stopped at the $10\,000$-th iteration. 
%\sstitle{Reproducibility Environment}
%Due to the page limitation, the technical report with more details of our work is available at: \url{https://arxiv.org/abs/2110.03536}. 
%Our experiments are implemented at NVIDIA Geforce GTX 1080 Ti Graphics Cards. 

%The PyTorch code is released at: \url{https://github.com/leibniz-future-lab/SoundPrototypeCriticism}. 

\subsection{End-to-end Comparison with SOTA Systems}
We compare our performance with those of all state-of-the-art (SOTA) approaches on the official test set. Please note that, the studies using a different test set are not comparable. In \autoref{tab:baselines}, our study is mainly compared with both hand-crafted features on classic machine learning classifiers~\cite{jakovljevic2017hidden,chambres2018automatic,serbes2017automated} and time-frequency representations on deep neural networks\cite{ma2019lungbrn,yang2020adventitious,ren2022prototype}.

Our approach outperforms all state-of-the-art methods on the test set when both AS and UAR are employed for evaluation. In particular, the CNN8 with dilation obtains $52.89\,\%$ AS, which is significantly ($p<0.05$ in a one-tailed z-test) better than the $50.37\,\%$ in~\cite{ren2022prototype}. Moreover, the ResNet with dilation and attention achieves $46.82\,\%$ UAR, which is significantly ($p<0.001$ in a one-tailed z-test) better than the $36.16\,\%$ UAR in~\cite{ren2022prototype}. Further, our best models have better performance on SE, which is quite important in clinical practice. Both of our best models have dilated convolutional kernels, indicating dilated kernels can improve performance of local max pooling. 

\begin{table}[!h]
\centering
\footnotesize
%\vspace{-10pt}
\caption{Classification performance [\%] comparison with the SOTA approaches on the test set.}
\label{tab:baselines}
\vspace{-10pt}
%%\scalebox{0.75}{
\begin{tabular}{lllll} 
\toprule
  & SE & SP & AS & UAR  \\ 
\midrule
MFCC-HMM-GMM~\cite{jakovljevic2017hidden}  &--&-- & 39.56 &--     \\
MFCC-Decision Tree~\cite{chambres2018automatic} & 20.81 & 78.05 & 49.43 &--      \\ 
STFT-Wavelet-SVM~\cite{serbes2017automated}  &--&--&49.86 &--      \\
STFT-Wavelet-BiResNet~\cite{ma2019lungbrn}  &31.12&69.20 &50.16 &--     \\
STFT-ResNet-Attention~\cite{yang2020adventitious}  &17.84&81.25&49.55 &-- \\  
LogMel-CNN8-Prototype~\cite{ren2022prototype}& 27.78 & 72.96 & 50.37 & 36.16 \\
\hline
\textbf{Ours (CNN8-Dilation)} & 35.85 & 69.92 & \textbf{52.89} & 40.26 \\
\textbf{Ours (ResNet-Dilation-Att)} & 51.83 & 50.22 & 51.02 & \textbf{46.82} \\
\bottomrule
\end{tabular}
%}
\vspace{-10pt}
\end{table}

%\sstitle{Setting}
%Similar to section 2.5 of \cite{chen2019looks} and section V.A of \cite{chong2020towards}
%\sstitle{Results and discussion}
%\autoref{tab:baselines}
%Our prototype learning approach is compared to the above SOTA systems in \autoref{tab:baselines}. 

\subsection{Ablation Study}

% \zr{I hope we use more models here (see \autoref{tab:ablation}). @Yi}
We evaluate the robustness of our prototype and criticism selection approach against various DNN models: CNN8 and ResNet.
The performance of dilation and attention on the two CNN models are compared in \autoref{tab:ablation}. When comparing the UAR and AS values inside the two types of models, the performance of dilation and attention is better than that of the models with local max pooling layers in some cases. When we compare the two types of models, ResNet mostly outperforms CNN8 probably due to the residual blocks in ResNet. Interestingly, high UAR values do not always lead to high AS values. We think this is highly related to the class-imbalance nature. 

\begin{table}[!h]
\centering
\footnotesize
%\vspace{-10pt}
\caption{The ablation study of the model performance [\%] on residual block, dilation, and attention. 
%The recall score on each class is shown for the four-class classification task.
}
\label{tab:ablation}
%\vspace{-10pt}
%\scalebox{0.75}{
\begin{tabular}{lp{.5cm}|p{.5cm}p{.5cm}|p{.5cm}p{.5cm}p{.5cm}}
\toprule
&\multicolumn{2}{c}{Four-class} & \multicolumn{4}{c}{Binary}\\
\cmidrule(lr){2-3}  \cmidrule(lr){4-7}
&Dev &Test &Dev &\multicolumn{3}{c}{Test} \\
&UAR & UAR & AS& SE & SP & AS  \\ 
\midrule
CNN8\cite{ren2022prototype} & --& 40.36& 52.99& 39.42 & 59.72 & 49.57\\
CNN8-Att & 38.51 & 42.75 & 49.56 & 43.76 & 49.65 & 46.70 \\
CNN8-Dila & 34.75 & 40.26 & 53.27 & 35.85 & 69.92 & \textbf{52.89} \\
CNN8-Dila-Att & 41.55 & 45.45 & 50.83 & 49.62 & 46.93 & 48.27 \\
ResNet & 41.69 & 45.33 & 54.48 & 43.67 & 58.01 & 50.84\\
ResNet-Att & 37.59 & 43.62 & 47.66 & 39.51 & 62.76 & 51.13 \\
ResNet-Dila & 37.20 & 43.39 & 52.65 & 46.73 & 44.59 & 45.66\\
ResNet-Dila-Att & 39.51 & \textbf{46.82} & 52.92 & 51.83 & 50.22 & 51.02\\
\bottomrule
\end{tabular}
%}
%\vspace{-10pt}
\end{table}

\subsection{Sensitivity Analysis}

% \zr{a figure analysing the number of iteration stepsin ifgsm}

\begin{figure}
    \centering
    \begin{subfigure}[b]{.5\textwidth}
        \centering
        \includegraphics[width=0.95\textwidth]{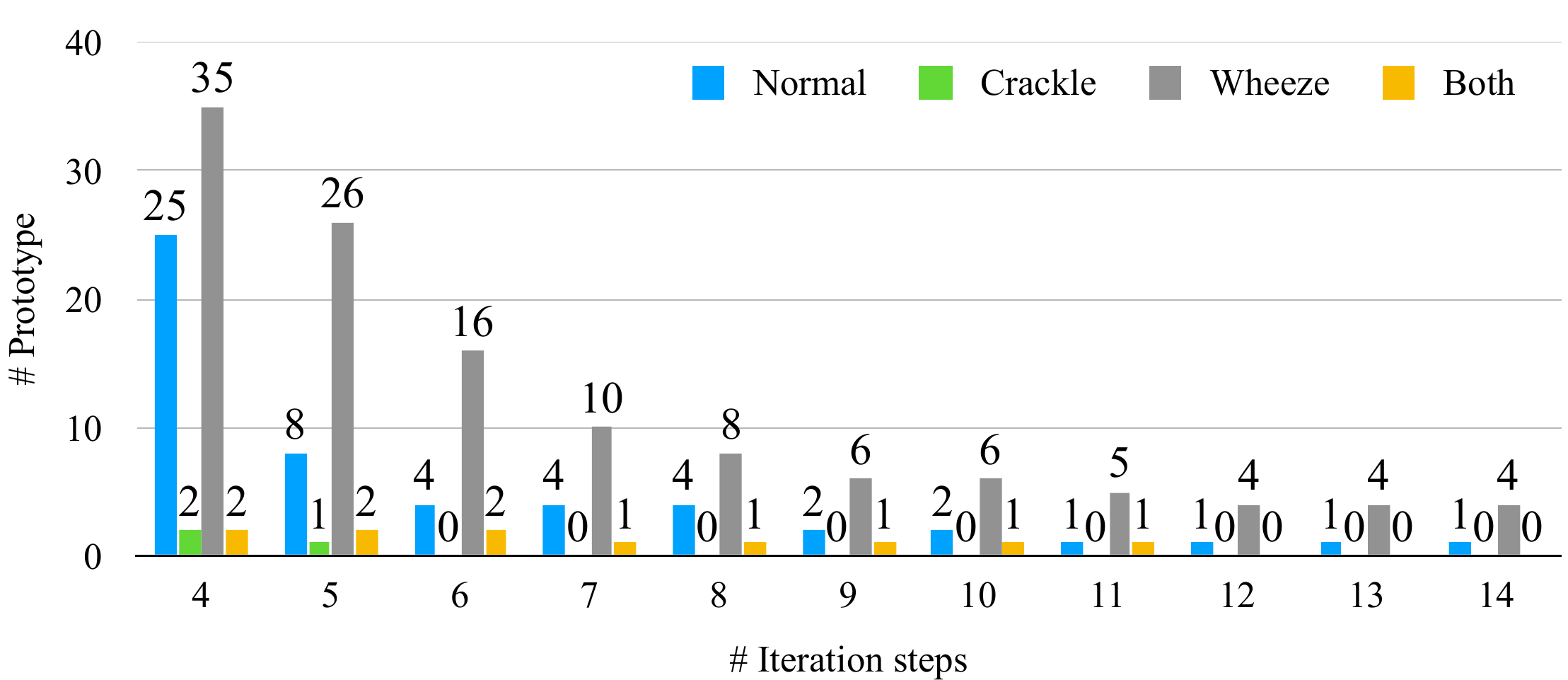}
%         \vspace{-5pt}
        \caption{Number of prototypes with different $I$.}
        \label{fig:proto_number}
%         \vspace{-5pt}
    \end{subfigure}
    \begin{subfigure}[b]{.5\textwidth}
        \centering
        \includegraphics[width=0.95\textwidth]{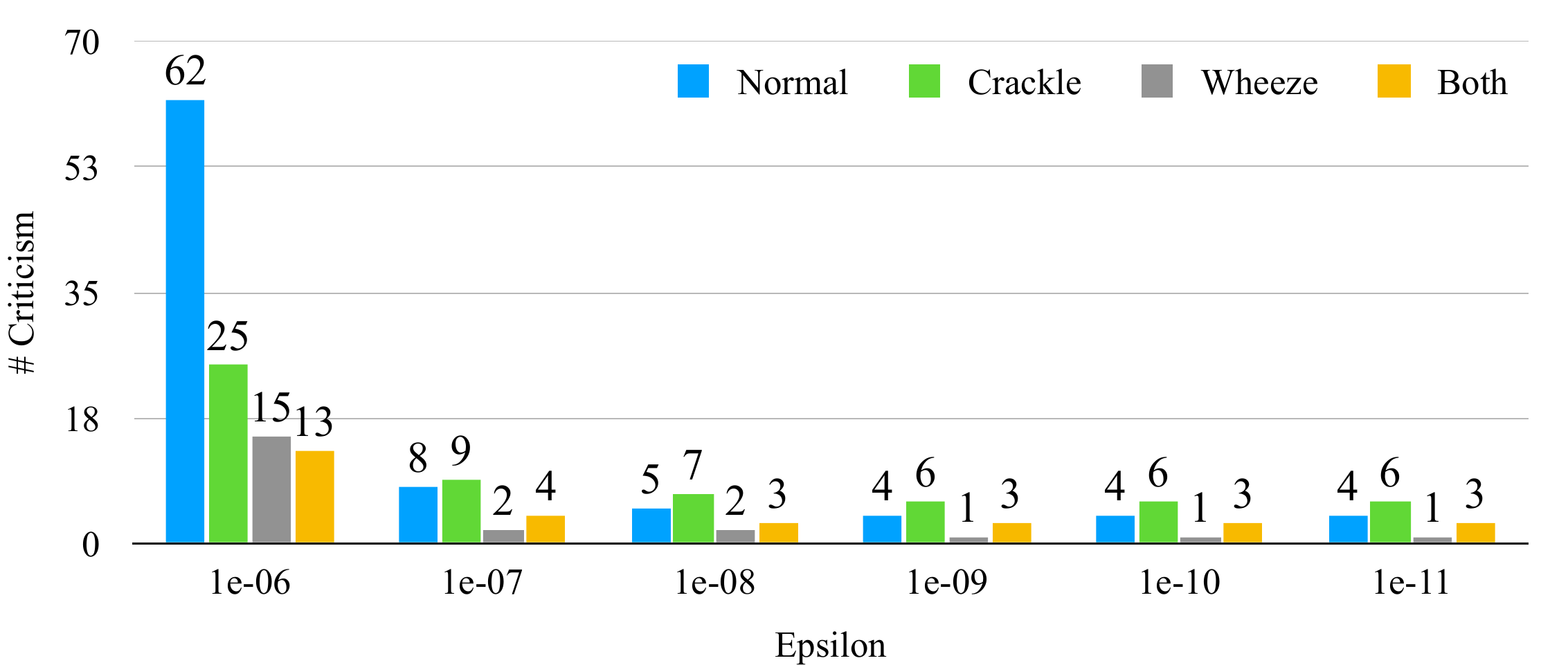} 
%        \vspace{-5pt}
        \caption{Number of criticisms with different $\epsilon$.}
        \label{fig:criti_number}
%        \vspace{-5pt}
    \end{subfigure}
    \caption{Analysis of the number of prototypes and criticisms.}
    \label{fig:proto_criti_number}
    \vspace{-15pt}
\end{figure}

We analyse the number of prototypes under different iteration steps of IFGSM and number of criticisms under different $\epsilon$ values when $I=1$ in \autoref{fig:proto_criti_number} based on the developed dilated ResNet with the attention mechanism. In \autoref{fig:proto_criti_number} (a), the number of prototypes for each class is decreasing when IFGSM keeps iterating, indicating the generated adversarial data is stronger with larger $I$ values. In each class, the number of criticisms is also decreasing when $\epsilon$ decreases (\ie adversarial data is becoming more similar to real data). Setting appropriate $I$ and $\epsilon$ is helpful for searching effective prototypes.

%\zr{t-sne figure}

% \begin{figure}[t]
%     \centering
%     \includegraphics[scale=0.40]{figures/t1.pdf}
%     \vspace{-10pt}
%     \caption{ The t-SNE figure based on intermediate features of $4\,142$ training samples. $*$ means selected prototypes and $+$ indicates the selected criticisms  
%     }
%     \label{fig:tsne}
% \end{figure}

% \edit{The number of layers? (the theory of IFGSM is, the deeper model is easier to attack)}

%The t-SNE figure we show here is based on the features after the $4$-th ResNet block.

\subsection{Visualisation of Prototypes and Criticisms}

\begin{figure}[!h]
    \centering
    \begin{subfigure}[b]{.48\textwidth}
    \includegraphics[trim={1.8cm 1.05cm 0 1.65cm},clip,width=.24\textwidth]{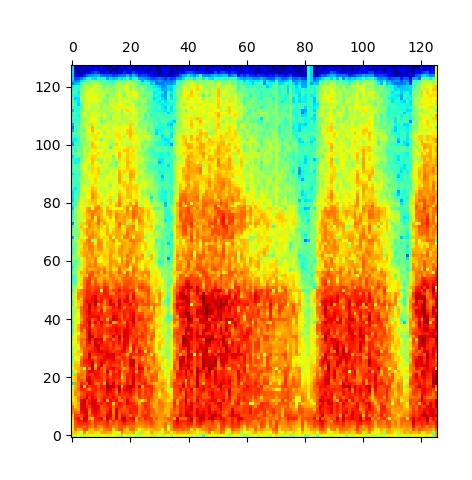}
    \includegraphics[trim={1.8cm 1.05cm 0 1.65cm},clip,width=.24\textwidth]{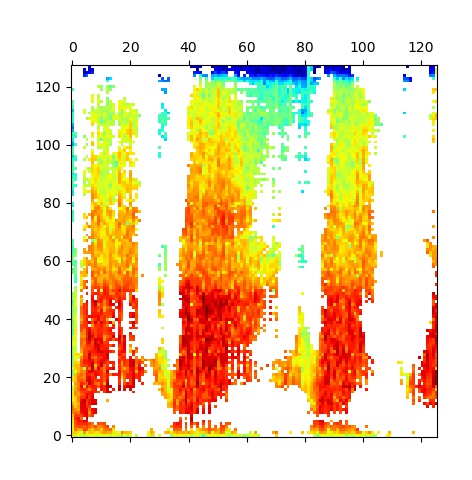}
    \includegraphics[trim={1.8cm 1.05cm 0 1.65cm},clip,width=.24\textwidth]{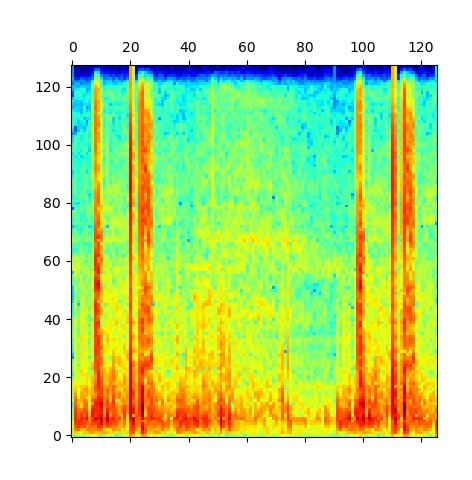}
    \includegraphics[trim={1.8cm 1.05cm 0 1.65cm},clip,width=.24\textwidth]{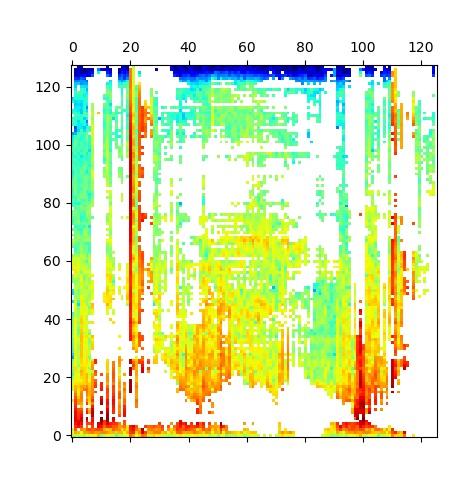}
    \label{fig:visulise_normal}
    \caption{Normal}
    \end{subfigure}
    \begin{subfigure}[b]{.48\textwidth}
    \includegraphics[trim={1.8cm 1.05cm 0 1.65cm},clip,width=.24\textwidth]{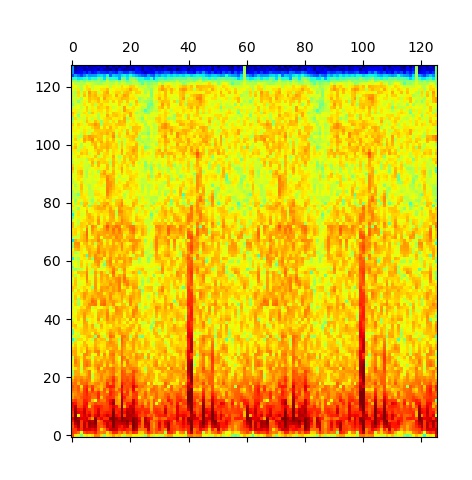}
    \includegraphics[trim={1.8cm 1.05cm 0 1.65cm},clip,width=.24\textwidth]{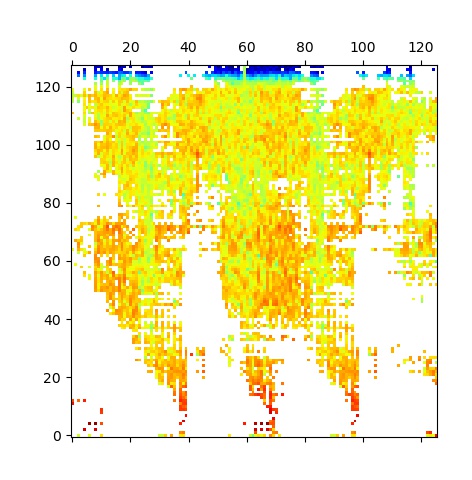} 
    \includegraphics[trim={1.8cm 1.05cm 0 1.65cm},clip,width=.24\textwidth]{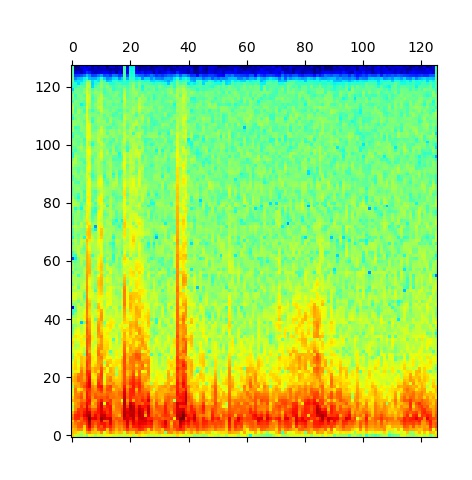}
    \includegraphics[trim={1.8cm 1.05cm 0 1.65cm},clip,width=.24\textwidth]{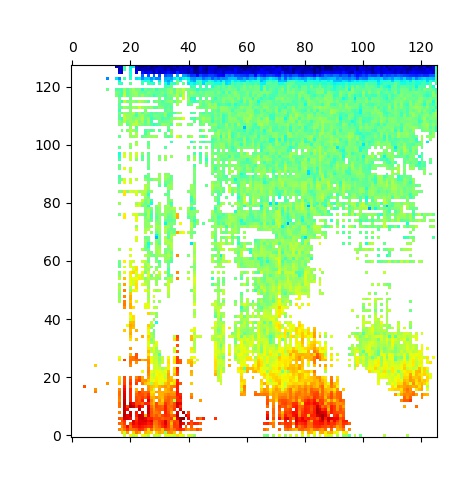}
    \label{fig:visulise_crackle}
    \caption{Crackle}
    \end{subfigure}
    \begin{subfigure}[b]{.48\textwidth}
    \includegraphics[trim={1.8cm 1.05cm 0 1.65cm},clip,width=.24\textwidth]{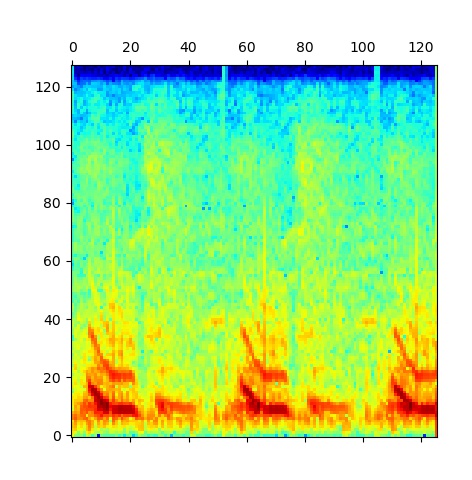}
    \includegraphics[trim={1.8cm 1.05cm 0 1.65cm},clip,width=.24\textwidth]{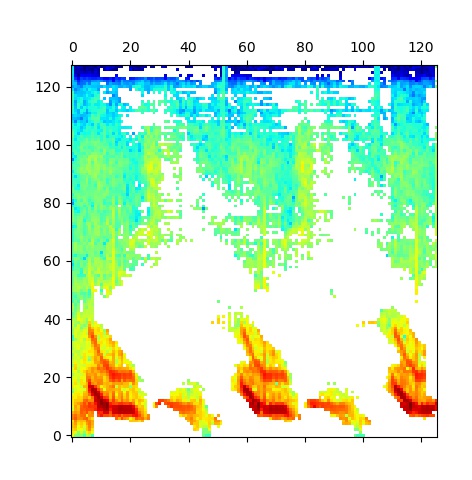}
    \includegraphics[trim={1.8cm 1.05cm 0 1.65cm},clip,width=.24\textwidth]{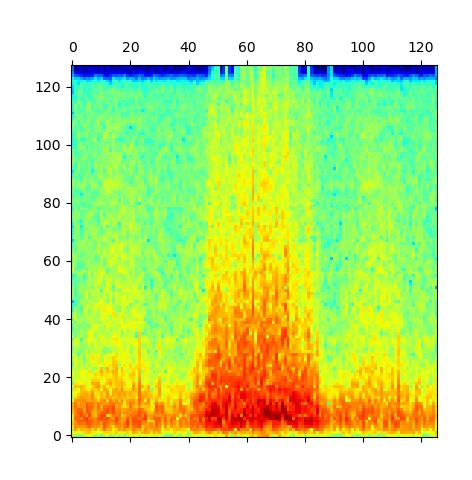}
    \includegraphics[trim={1.8cm 1.05cm 0 1.65cm},clip,width=.24\textwidth]{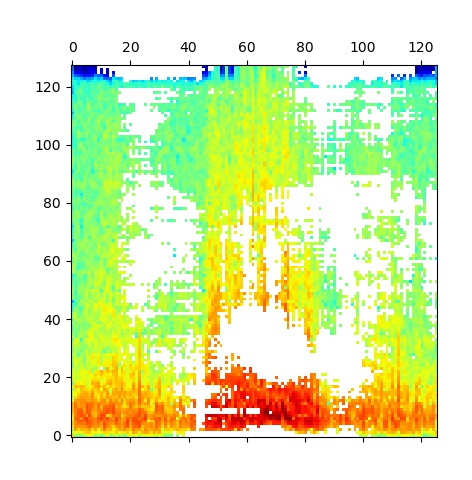}
    \label{fig:visulise_wheeze}
    \caption{Wheeze}
    \end{subfigure}    
    \begin{subfigure}[b]{.48\textwidth}
    \includegraphics[trim={1.8cm 1.05cm 0 1.65cm},clip,width=.24\textwidth]{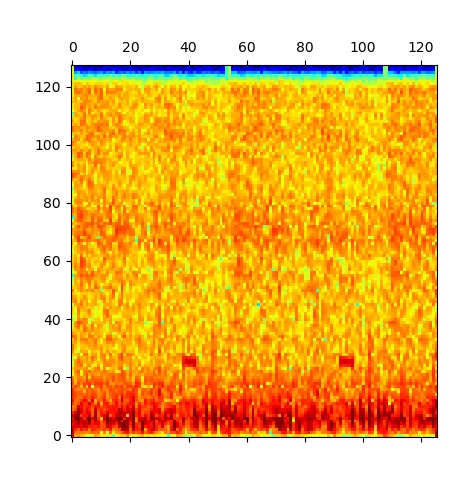}
    \includegraphics[trim={1.8cm 1.05cm 0 1.65cm},clip,width=.24\textwidth]{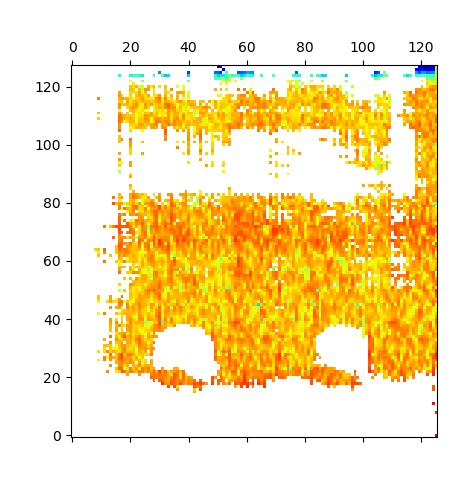}
    \includegraphics[trim={1.8cm 1.05cm 0 1.65cm},clip,width=.24\textwidth]{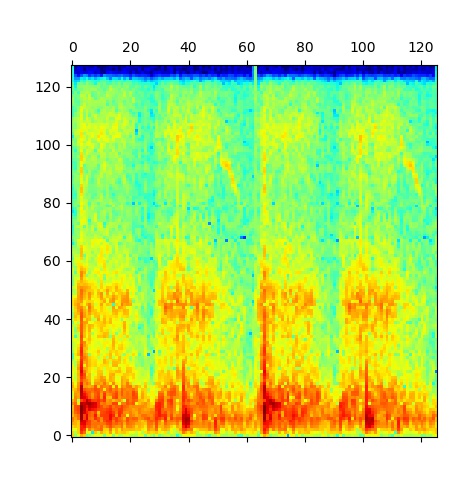}
    \includegraphics[trim={1.8cm 1.05cm 0 1.65cm},clip,width=.24\textwidth]{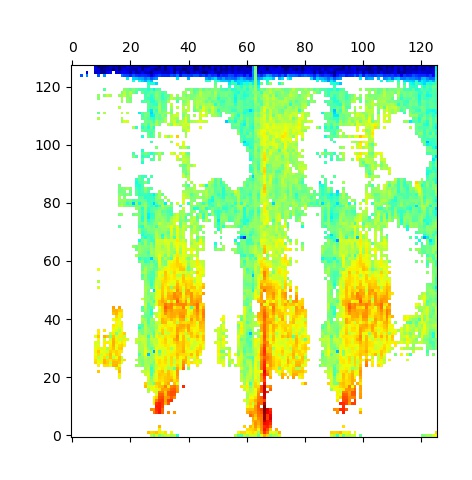}
    \label{fig:visulise_both}
    \caption{Both}
    \end{subfigure}
%     \vspace{-20pt}
    \caption{Visualisation of prototypes and criticisms, as well as their contribution parts for respiratory sound classification. X-axis: Time steps, Y-axis: Mel frequency bins. The first column contains the prototypes, the second column shows the contribution parts of prototypes, the third shows the criticisms, and the final one shows the criticisms' contribution parts. }
    \label{fig:visualisation}
    \vspace{-15pt}
\end{figure}

The log Mel spectrograms of the selected prototypes and criticisms are depicted in \autoref{fig:visualisation}. The four prototypes are representive sounds in the four classes. Particularly, the normal prototype sound is regular breathing in \autoref{fig:visualisation} (a). As the crackle sounds are explosive, short-duration transient sounds, they can have a big range of magnitude and frequency content \cite{quintas2013multi}. The selected prototype has the consistent characteristics on the duration and frequency \autoref{fig:visualisation} (b). Compared to crackle sounds, wheezes have relatively long duration \cite{quintas2013multi}. We can see the wheeze prototype has longer duration than the crackle one for each respiratory cycle \autoref{fig:visualisation} (c). The ``both'' class (\autoref{fig:visualisation} (d)) is a combination of crackle and wheeze, therefore, we can only see it is different from the normal one.

When comparing criticisms and prototypes, we can see that the typical characteristics of the sounds only appear in part of of the whole waveform, especially normal, crackle, and wheeze. We think this is also the reason that the criticisms are easy to be misclassified in a single iteration step of IFGSM.
We also project the attention heat maps to the prototypes and criticisms with a threshold at the attentions tensor' middle values. The time-frequency bins of a prototype/criticism are visualised when the corresponding bin in the attention heat map is larger than the threshold.
We can see that, the respiratory cycles are preserved in the projections of normal and wheeze prototype sounds. In the projection of the crackle prototype, the non-respiratory part is reserved, probably because the respiratory duration is too short. For the projection of criticisms, the respiratory parts in all four classes are highlighted. Different from the crackle prototype, the non-respiratory part is learnt with low coeffients in the attention heat-map of criticism. We think this is caused by fewer high frequency sounds in the crackle criticism.

% normal: 
% pro: good, heart sound
% cri: heart sound, speech

% crackle
% pro: difficult to breath, heart sound
% cri: noise, not clear

% wheeze
% pro: hu xi ji cu, weak
% cri: noise or heart sound bigger than breath

% both
% pro: hu xi ji cu, difficult
% cri: like normal, but weak

\section{Conclusion}

Existing explainable classification methods do not often consider bias in data. This paper developed a unified example-based explanation for respiratory sound classification by selecting prototypes and criticisms via an iterative fast gradient sign method. Not only applicable for any deep neural networks, our explanations can assist physicians in exploring extreme cases and making informed decisions. Experiments show that our approach can outperform the baselines, and achieve average score of 52.89\,\% and unweighted average recall of 46.82\,\%.
%uncover XX\% of model mistakes and can achieve nearly XX\% of the original model performance using only a nearest prototype classifier.
In future work, we will explore the affect of adversarial attacks by analysing the attention map of adversarial data. We also plan to explore other types of explanations such as counterfactuals~\cite{goyal2019counterfactual}.

%\section{Acknowledgements}

%This research was partially funded by the Federal Ministry of Education and Research (BMBF), Germany under the project LeibnizKILabor with grant No. 01DD20003.

\balance
\bibliographystyle{IEEEtran}

\bibliography{refs}

% \begin{thebibliography}{9}
% \bibitem[1]{Davis80-COP}
%   S.\ B.\ Davis and P.\ Mermelstein,
%   ``Comparison of parametric representation for monosyllabic word recognition in continuously spoken sentences,''
%   \textit{IEEE Transactions on Acoustics, Speech and Signal Processing}, vol.~28, no.~4, pp.~357--366, 1980.
% \bibitem[2]{Rabiner89-ATO}
%   L.\ R.\ Rabiner,
%   ``A tutorial on hidden Markov models and selected applications in speech recognition,''
%   \textit{Proceedings of the IEEE}, vol.~77, no.~2, pp.~257-286, 1989.
% \bibitem[3]{Hastie09-TEO}
%   T.\ Hastie, R.\ Tibshirani, and J.\ Friedman,
%   \textit{The Elements of Statistical Learning -- Data Mining, Inference, and Prediction}.
%   New York: Springer, 2009.
% \bibitem[4]{YourName17-XXX}
%   F.\ Lastname1, F.\ Lastname2, and F.\ Lastname3,
%   ``Title of your INTERSPEECH 2022 publication,''
%   in \textit{Interspeech 2022 -- 23\textsuperscript{rd} Annual Conference of the International Speech Communication Association, September 18-22, Incheon, Korea, Proceedings, Proceedings}, 2022, pp.~100--104.
% \end{thebibliography}

\end{document}